\documentclass[aps,prb,twocolumn,superscriptaddress,showpacs]{revtex4}
\usepackage{graphicx}
\usepackage{amssymb}
\usepackage{amsfonts}
\usepackage{bm}
\begin{document}

\title{Origin of ferromagnetism in (Zn,Co)O from magnetization and spin-dependent magnetoresistance}

\author{T. Dietl}
\affiliation{Institute of Physics, Polish Academy of Science,
al.~Lotnik\'ow 32/46, PL 02-668 Warszawa, Poland}
\affiliation{Institute of Theoretical Physics, Warsaw University,
PL 00-681 Warszawa, Poland} \affiliation{ERATO Semiconductor
Spintronics Project, Japan Science and Technology Agency,
al.~Lotnik\'ow 32/46, PL 02-668 Warszawa, Poland}
\author{T. Andrearczyk}
\affiliation{Institute of Physics,
Polish Academy of Science, al.~Lotnik\'ow 32/46, PL 02-668 Warszawa,
Poland}
\author{A. Lipi\'nska}
\affiliation{Institute of Physics,
Polish Academy of Science, al.~Lotnik\'ow 32/46, PL 02-668 Warszawa,
Poland}
\author{M. Kiecana}
\affiliation{Institute of Physics,
Polish Academy of Science, al.~Lotnik\'ow 32/46, PL 02-668 Warszawa,
Poland}
\author{Maureen Tay}
\affiliation{Department of Electrical and Computer Engineering,
National University of Singapore, 4 Engineering Drive 3, Singapore 117576}
\affiliation{Data Storage Institute, 5 Engineering Drive 1, Singapore 117608}
\author{Yihong Wu} \affiliation{Department of Electrical and Computer Engineering,
National University of Singapore, 4 Engineering Drive 3, Singapore 117576}

\date{\today}

\begin{abstract}
In order to elucidate the nature of ferromagnetic signatures
observed in (Zn,Co)O we have examined experimentally and theoretically
magnetic properties and spin-dependent quantum localization effects
that control low-temperature magnetoresistance.
Our findings, together with a through structural characterization,
substantiate the model assigning spontaneous magnetization
of (Zn,Co)O to uncompensated spins at the surface of antiferromagnetic
nanocrystal of Co-rich wurtzite (Zn,Co)O. The model explains a large
anisotropy observed in both magnetization
and magnetoresistance in terms of spin hamiltonian of Co ions in the
crystal field of the wurtzite lattice.
\end{abstract}


\maketitle

\section{Introduction}
Since the discovery of ferromagnetic signatures in (Zn,Co)O,\cite{Ueda:2001}
this material has reached the status of a model system representing
the ever increasing class of wide-band gap diluted magnetic semiconductors (DMS),
diluted magnetic oxides (DMO), and non-magnetic
oxides, in which challenging high-temperature ferromagnetism is observed
under some growth conditions and/or post-growth processing.\cite{Rode:2003,Kolesnik:2004,Venkatesan:2004,Chambers:2006}
In this paper, we present results
of magnetization and magnetoresistance measurements of (Zn,Co)O:Al, whose
quantitative interpretation substantiates the recent suggestion\cite{Dietl:2006_a,Dietl:2006_b}
that the puzzling ferromagnetic behavior can originate from uncompensated
spins at the surface of wurtzite nanocrystals of Co-rich (Zn,Co)O.
These nanocrystals are immersed in a Co-poor paramagnetic (Zn,Co)O matrix which,
as we demonstrate here, determines magnetotransport properties of the system. Since the nanocrystals are coherent, {\em i.~e.}, their crystallographic structure and lattice constant are identical
to the surrounding (Zn,Co)O, they escape from the detection
by standard HR XDR or TEM.

\section{Samples and experimental setup}

The studied (Zn,Co)O thin films are deposited by
sputtering on (0001) sapphire ($\alpha$-Al$_2$O$_3$) substrates,
so that the $c$-axis of the wurtzite structure is perpendicular to
the film plane. A multi-chamber high vacuum system with a base
pressure~$< 10^{-7}$~Torr is employed. Sintered ZnO,
Al$_2$O$_3$, and Co targets are used as the sputtering sources for
ZnO, Al, and Co, respectively. The samples are sputtered in an
atmosphere of pure Ar gas at a pressure of 5~mTorr. Prior to
deposition, the substrates are cleaned using Ar reverse
sputtering at 20~mTorr in a pre-cleaning chamber. A series of
experiments has been carried out to optimize the substrate
temperature and sputtering powers in order to obtain films with
low resistivity. At a substrate temperature of $500^\circ$C, ZnO
sputtering power of 150~W and Al$_2$O$_3$ sputtering power of
30~W, the resistivity of as-grown ZnO:Al film was about
$10^{-3}$~$\Omega$cm at room temperature. These conditions are
employed to deposit Co-doped samples in which the Co
composition $x$ varied by changing the Co sputtering power
from 3 to 50~W. The obtained x-ray diffraction patterns and the
corresponding pole figure diagrams show that the films are
well textured in the (0001) direction for $x < 0.25$.

We focus here on results obtained for non-magnetic ZnO:Al and
for Zn$_{0.95}$Co$_{0.05}$O:Al showing clear ferromagnetic signatures.
The Co content $x$ is determined by
x-ray photoelectron spectroscopy. The thickness for both films
$d=200$~nm has been confirmed by transmission electron microscopy (TEM).
A careful examination by high resolution x-ray diffraction (HR XDR)  and HR TEM
demonstrates a good wurtzite
structure of the films without evidences for precipitates of
a secondary crystallographic phase, such as Co precipitates which
are often present, especially in samples with high Co content.

We employ a high-field SQUID system for the magnetic studies.
For electrical characterizations, Hall bars with a size $324\times80$~$\mu$m$^2$
are fabricated using a direct
laser writer in a combination with ion milling. An Al/Au bilayer is
deposited in order to obtain Ohmic contacts. All
electrical measurements have been carried by a DC method  with the current
set as $100$~$\mu A$ -- a low limit ensuring an adequate signal-to-noise ratio in the
employed setup. No indications of anomalous Hall effect have been
detected. A uniform Al doping results in the high and
almost temperature independent electron concentrations $n =1.7$ and
$1.1\times 10^{20}$~cm$^{-3}$, for $x=0$ and 5\%, respectively,
according to the Hall effect data. The presence of degenerate
electrons with a band mobility value indicates that the Co acceptor
state Co$^{2+}$/Co$^{+1}$ is well above the bottom of the conduction band in ZnO.

\section{Experimental results}
\subsection{Magnetization measurements}

\begin{figure}
\includegraphics[]{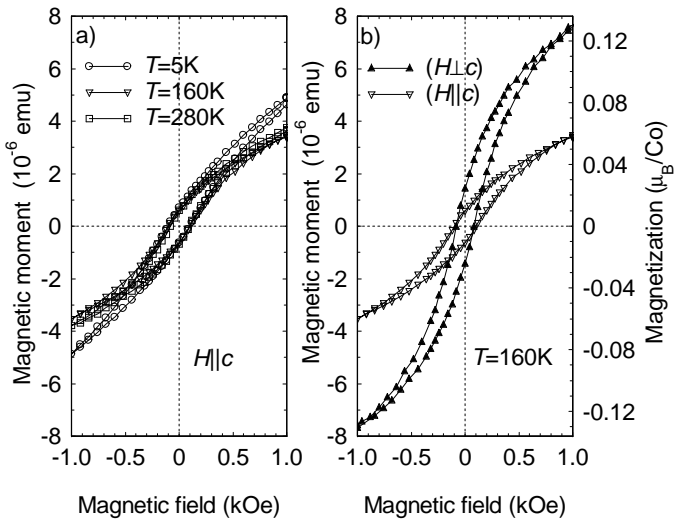}
\caption{Hysteresis loop of
Zn$_{0.95}$Co$_{0.05}$O:Al measured as a function of external
magnetic field; (a) for different temperatures and (b) for two
directions of the magnetic field in respect to the wurtzite
\emph{c}-axis.  A temperature independent contribution linear in the field is subtracted
from the data.} \label{fig:Magn1}
\end{figure}

Magnetization measurements of the Zn$_{0.95}$Co$_{0.05}$O:Al
sample show well-developed hysteresis up to room temperature with
a coercivity of about 50~Oe, as shown in Fig.~\ref{fig:Magn1}.
The presence of a robust ferromagnetism is indeed surprising
since the {\em electron}-mediated spin-spin interaction is not
expected to result in a ferromagnetic order above 1~K in DMS in
question.\cite{Dietl:1997,Andrearczyk:2001,Spaldin:2004} An important
aspect of the results is a strong anisotropy: magnetization is
significantly larger for the in-plane magnetic field. This
indicates that shape magnetic anisotropy is overcompensated by crystalline
magnetic anisotropy. The data point to the presence of the
easy plane perpendicular
to the $c$-axis of the wurtzite structure.

\begin{figure}
\includegraphics[]{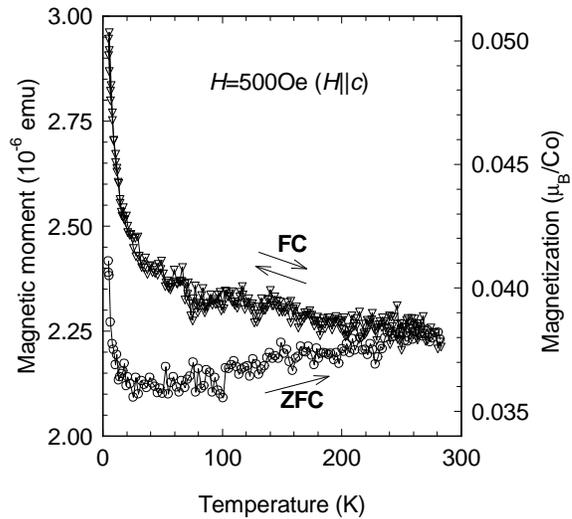}
\caption{Magnetization of Zn$_{0.95}$Co$_{0.05}$O:Al measured as a
function of temperature, at magnetic field 500~Oe in $H \parallel
c$ configuration. The arrows indicate direction of temperature
ramp. The zero field cooled (ZFC) and field cooled (FC)
data were obtained after earlier
cooling of the sample from 280~K down to 5~K without and with the applied
magnetic field, respectively. A temperature independent contribution linear in the field is subtracted
from the data.} \label{fig:Magn2}
\end{figure}

A comparison of magnetization data obtained after cooling without the magnetic field
and with the magnetic field depicted in Fig.~\ref{fig:Magn2}
shows a behavior indicative of superparamagnetism. According to these
results, the blocking reaches the room temperature.

\begin{figure}
\includegraphics[]{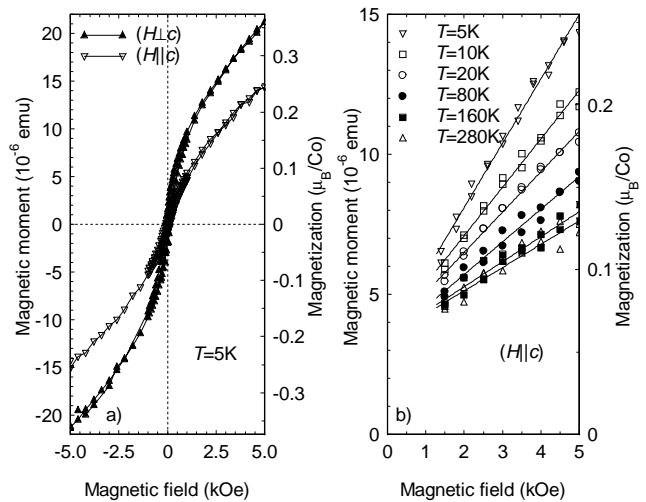}
\caption{Magnetization of Zn$_{0.95}$Co$_{0.05}$O:Al measured as a
function of the external magnetic field; (a) for two directions of
the magnetic field in respect to the wurtzite \emph{c}-axis;  (b)
for different temperatures. The straight lines result from fitting
to the data in a high
field range. A temperature independent contribution linear in the field is subtracted
from the data.}
\label{fig:Magn3}
\end{figure}

As shown in in Fig.~\ref{fig:Magn3}, the hysteresis are superimposed on a
background which is linear in the magnetic field with a slop
increasing when temperature decreases. We evaluate
that about 50\% of Co ions contributes to this paramagnetic response.

\subsection{Magnetoresitance: non-magnetic \textit{n}-ZnO:Al}

\begin{figure}
\includegraphics[]{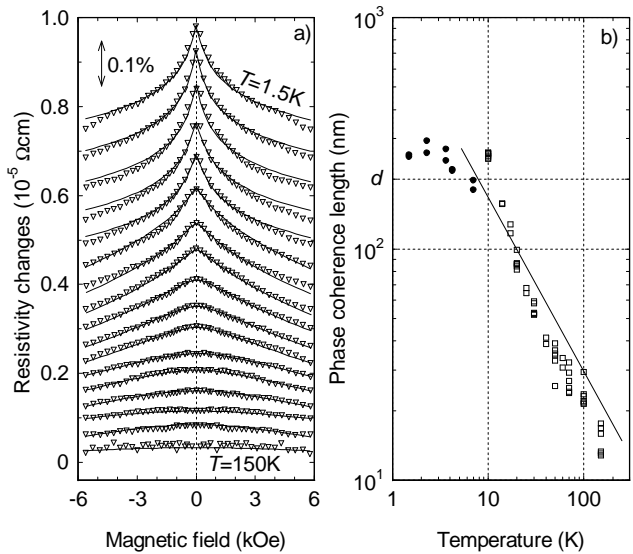}
\caption{(a) Resistivity changes in the magnetic field ($H
\parallel c$) for ZnO:Al (points) measured at $T=1.5$, 2.3,
3.6, 4.2, 7, 10, 14, 17, 20, 25, 30, 40, 50, 60, 70, 100 and
150~K. Curves are vertically shifted for clarity. The fitted
curves (lines) are obtained within the weak localization (WL)
theory. (b) Phase coherence length $L_{\varphi}$ (which is the
fitting parameter) as a function of temperature. The dots denote
$L_{\varphi}$ values determined by fitting 2D WL theory\cite{Altshuler:1982,Kawabata:1980,Altshuler:1985}
and the squares by 3D one ($L_{\varphi} < d$).\cite{Altshuler:1982,Altshuler:1985}
Straight line shows the $T^{-3/4}$ dependence expected for 3D
disordered systems.} \label{fig:2}
\end{figure}

Figure~\ref{fig:2}(a) shows magnetoresistance (MR) of ZnO:Al in the magnetic field $H$
applied parallel to the $c$ axis. We have checked and confirmed
that the shape of MR curve is independent of the applied field
direction. As seen, MR is negative and becomes stronger with
decreasing temperature. Such a character and magnitude of MR (of
the order of 0.1\% at $T=1.5$~K and $H= 5$~kOe) are similar to
those observed previously for ZnO:Al films grown by pulsed laser
deposition technique\cite{Andrearczyk:2005} and for an
accumulation layer on ZnO.\cite{Goldenblum:1999} We anticipate,
therefore, that similarly to those cases, MR results form quantum
corrections to conductivity of disordered systems, that is from a
destructive effect of the magnetic field (vector potential) on
constructive interference corresponding to two time-reversal paths
along the same self-crossing trajectories. The influence of this
effect on  MR can be quantitatively described in the weakly
localized regime $k_Fl \gg 1$, where $k_F$ is the Fermi wave
vector and $l$ is the mean free path.\cite{Altshuler:1985} The value $k_Fl=7$
is determined for our ZnO:Al sample.

In order to evaluate MR theoretically, we follow the procedure employed
previously,\cite{Andrearczyk:2005}
treating the phase coherence length $L_{\varphi}$ as the only
fitting parameter to independently determined MR curves at particular
temperatures. The determined values of $L_{\varphi}$
are employed in the MR simulations for (Zn,Co)O described below.
The magnitude of $L_{\varphi}$ and its temperature dependence $L_{\varphi} \sim T^{-3/4}$
are similar to those found previously,\cite{Andrearczyk:2005} though
$L_{\varphi}(T)$ determined here starts to saturate at higher temperature.
We associate this difference to electron heating by a relatively high current
applied in the present setup.

\subsection{Magnetoresitance: magnetic
 \textit{n}-Zn$_{0.95}$Co$_{0.05}$O:Al }

\begin{figure}
\includegraphics[]{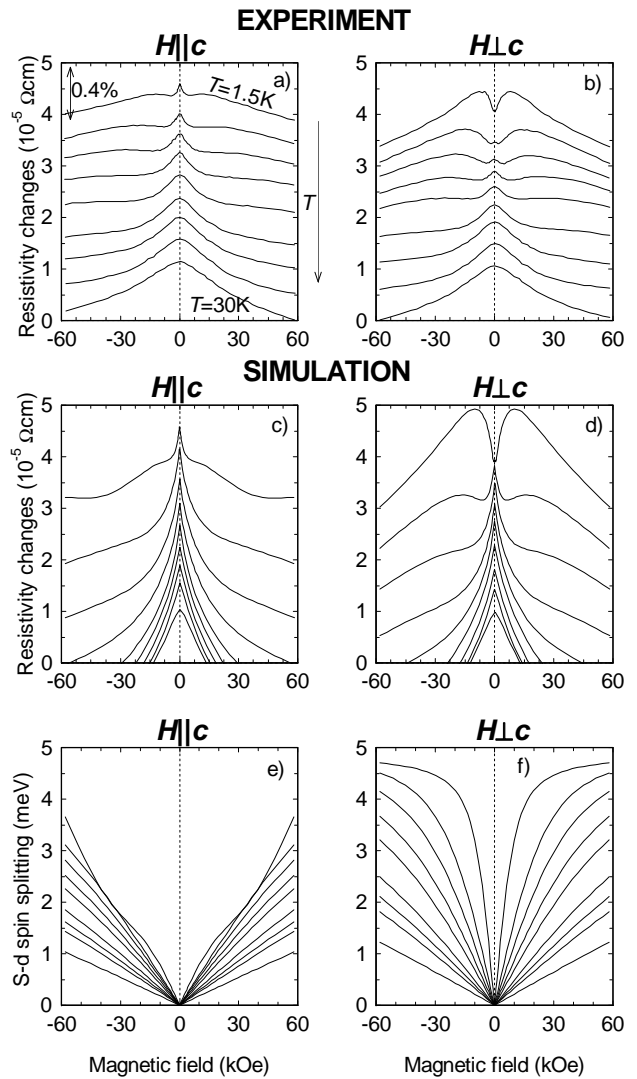}
\caption{Resistivity changes of Zn$_{0.95}$Co$_{0.05}$O:Al in the
magnetic field, measured (a,b) and simulated (c,d) for
temperatures $T=1.5$, 4.2, 6, 8, 10, 14, 17, 20 and 30~K. Curves
are vertically shifted for clarity. Simulation of magnetic field
dependence of Co-related spin-splitting (e,f) are performed for
the same set of temperatures. The configuration of the external
magnetic field is $H\parallel c$ for (a,c,e) and $H\perp c$ for
(b,d,f).} \label{fig:MRZnCoO}
\end{figure}

The resistivity changes in the magnetic field as found for
Zn$_{0.95}$Co$_{0.05}$O:Al are shown in Figs.~\ref{fig:MRZnCoO}(a) and
\ref{fig:MRZnCoO}(b) for $H\parallel c$ and $H\perp c$, respectively. A
positive MR component, absent in \textit{n}-ZnO, is seen to take
over with lowering temperature. A similar behavior, {\em i.~e.},
the appearance of temperature-dependent {\em positive} MR in the
presence of magnetic ions, has been previously observed for
\textit{n}-(Cd,Mn)Se,\cite{Sawicki:1986}
\textit{n}-(Cd,Mn)Te,\cite{Shapira:1990,Jaroszynski:2007}
\textit{n}-(Cd,Zn,Mn)Se,\cite{Smorchkova:1997}, and \textit{n}-(Zn,Mn)O.\cite{Andrearczyk:2005}
Remarkably, however, no spontaneous
magnetization was revealed in those materials,
and positive MR was quantitatively
described\cite{Andrearczyk:2005,Sawicki:1986,Shapira:1990,Jaroszynski:2007,Smorchkova:1997}
by the effect of the field-induced
giant spin-splitting on disorder-modified electron-electron
interactions.\cite{Altshuler:1985} In
accord with such an interpretation, no corresponding MR was found
in ferromagnetic (Ga,Mn)As,\cite{Matsukura:2004} where hole
states are spin-polarized already in the absence of an external
magnetic field. On the other hand, contrary to the previous
studies of paramagnetic wurtzite (wz) DMS,\cite{Andrearczyk:2005,Sawicki:1986}
MR of (Zn,Co)O reveals a
dependence on the direction of the magnetic field in respect to
the $c$-axis. As visible in Fig.~\ref{fig:MRZnCoO}, the positive
component is stronger for the $H\perp c$ case. We have checked
that this MR does not depend on the orientation of the magnetic
field in respect to the current direction. A sensible model of
(Zn,Co)O has to elucidate, therefore, why {\em no} ferromagnetic
signatures are observed in electron transport as well as explain
large anisotropy of MR.

\section{Discussion}

We will now demonstrate that our findings, together with a number of puzzling data accumulated
over the recent years,\cite{Rode:2003,Kolesnik:2004,Venkatesan:2004,Chambers:2006,Kobayashi:2005,Sati:2006}
can readily be interpreted in terms of nano-scale spinodal decomposition
into antiferromagnetic  Co-rich wz-(Zn,Co)O nanocrystals and a matrix of wz-(Zn,Co)O accounting
for the paramagnetic contribution.
According to this model, the ferromagnetic signatures result from
uncompensated spins at the nanocrystal surface\cite{Dietl:2006_b}
and will be visible below both the N\'eel temperature $T_{\mathrm{N}}$
and the blocking temperature $T_{\mathrm{B}}$ of the nanocrystals.
Such an effect has actually been visible in the case of NiO nanoparticles\cite{Winkler:2005}
and analyzed theoretically.\cite{Eftaxias:2005}

The  quantitative examination  of magnetic susceptibility as a
function of $x$ in paramagnetic wz-Zn$_{1-x}$Co$_{x}$O
led to the extrapolated antiferromagnetic Curie-Weiss temperature
as high as $|\Theta| = 950\pm 100$~K for $x = 1$,\cite{Kolesnik:2004}
indicating that, indeed,
$T_{\mathrm{N}}$ of the Co-rich wz-(Zn,Co)O can well surpass the room temperature.
Furthermore, a non-zero value
of the orbital momentum in the case of the  $S = 3/2$ spins of Co$^{2+}$ ions
in ZnO results in a relatively large magnitude
of the single-ion magnetic uniaxial anisotropy energy $D = 3.97$~K
with the easy plane perpendicular to the $c$-axis.\cite{Sati:2006} Assuming that the in-plane
anisotropy is $10$ times smaller, the nanocrystals with the radius $r = 3.7$~nm
will exhibit $T_{\mathrm{B}} \approx 4\pi N_0r^3DS^2/(30\times 25)$ as high as $300$~K,
where $N_0 = 4.2\times 10^{22}$~cm$^{-3}$ is the cation concentration in ZnO.

\begin{figure}
\includegraphics[scale=1.0]{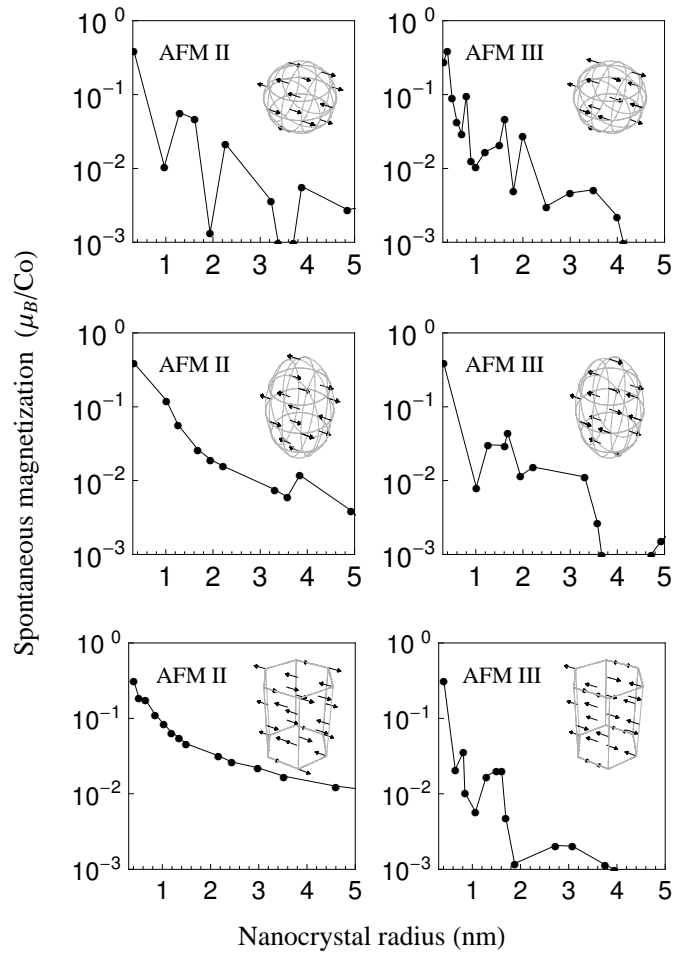}
\caption{Saturation magnetization of uncompensated Co-spins at
the surfaces of antiferromagnetic wurtzite CoO nanocrystals, calculated as a function of
the nanocrystal size for different nanocrystal shapes as indicated in the legend.
 It is assumed that a half (2.5\%) of Co atoms in
Zn$_{0.95}$Co$_{0.05}$O forms identical nanocrystals exhibiting type II or III
aniferromagnetic ordering.
The quoted radiuses correspond to spheres of the same volume.
The magnetization value is given in the Bohr magnetons
per one cobalt ion in Zn$_{0.95}$Co$_{0.05}$O.} \label{fig:Magn5}
\end{figure}

Figure~\ref{fig:Magn5} shows spontaneous magnetization computed according to the above model for various nanocrystal shapes and dimensions. We have investigated type II and type III antiferromagnetic (AF) spin arrangement in the wurtzite structure. \cite{Hines:1997}  A spherical nanocrystal of radius $r$ is defined by fixing the origin at a certain lattice site and including all spins within the distance $r$.\cite{Trohidou:1990} The ellipsoid and hexagonal nanocrystals have been constructed in a similar way, defining $r$ as the radius of a sphere of the same volume as the ellipsoid and hexagonal prism, respectively.  We see that for the wurtzite prism  and ellipsoid AFM II nanocrystals spontaneous magnetization attains significant values. In particular, for $r = 3.7$~nm, as quoted above,  spontaneous magnetization per Co ion reaches a few percent of the Bohr magneton, in agreement with the experimental results displayed in Figs.~1--3.

The presence of antiferromagnetic wz nanocrystals invoked here
has important consequences for the  interpretation of magnetization in thin (Zn,Co)O films.
In particular, the term linear in the magnetic field, so far assumed to result from substrate
diamagnetism and the film paramagnetism, can actually contain a sizable contribution
from the antiferromagnetic portion of the film. Furthermore, Co-precipitates are known
to act as nucleation centers for CoO,
which exert an exchange bias on the Co cores.\cite{Tracy:2005} Accordingly, the ferromagnetic-like
response may originate, in general, from uncompensated spins at the nanocrystal
surface but also from Co precipitates exchange biased by the surrounding Co-rich (Zn,Co)O.
In either case, the large magnitude of uniaxial crystal-field anisotropy of Co in ZnO, corresponding
to the anisotropy filed $H_{\mathrm{A}} = 53$~kOe,\cite{Sati:2006} elucidates
why the magnetic response observed by us (Figs.~1 and 3) and others
\cite{Venkatesan:2004} is much stronger for the $H\perp c$ configuration
comparing to the $H\parallel c$ case. At the same time, it
rules out precipitates of free standing metallic Co precipitates as the dominant source of ferromagnetic-like
response in these samples. Importantly, this view is supported by x-ray absorption,
x-ray magnetic circular dichroism (XMCD),
and photoemission experiments for ferromagnetic (Zn,Co)O,\cite{Kobayashi:2005}
which demonstrate that Co atoms substitute Zn, assume the high spin 2+ charge state, and
interact antiferromagnetically. Furthermore, within this scenario, the dependence of ferromagnetic
response on growth conditions and processing\cite{Chambers:2006,Venkatesan:2004}
results from a sensitivity of the nanocrystal aggregation process to the co-doping.\cite{Dietl:2006_a,Kuroda:2007}
The (Zn,Co)O nanocrystals in question can be, presumably, classified as
Mott-Hubbard antiferromagnets. The corresponding insulating character together
with a small magnitude of nanocrystal magnetization, account for weak ferromagnetic signatures
in MCD spectra near the band gap of the host.\cite{Chambers:2006,Ando:2006} In the same way
we explain the established here lack of ferromagnetic signatures in electron transport.

Taking the above arguments into account we assume that MR is
determined by the properties of the paramagnetic (Zn,Co)O matrix.
Owing to a large value of the relevant parameter $k_Fl\approx 3$
we can safely apply the previously developed approach\cite{Sawicki:1986,Andrearczyk:2005} to the weak localization
theory with the effects of electron-electron interactions taken
into account.\cite{Altshuler:1985} We
compute MR values employing the 3D formulae, as the diffusion
constant is smaller in (Zn,Co)O comparing to ZnO, so that the
dimensional cross-over occurs at lower temperature. There are two
parameters describing the effect: the magnitude of the interaction
in the triplet channel taken as\cite{Altshuler:1985,Sawicki:1986}
 $F_{\sigma} \equiv 2g_3=1$ and the
spin-splitting of the conduction band. The splitting contains the
Zeeman term $g^*\mu_BB$, with $g^*=2.0$, and the s-d contribution,
$-x_{\mathrm{eff}} \alpha N_0 \langle S_z \rangle$. The exchange
energy $\alpha N_0 =0.18$~eV is assumed, which was
determined for wz-(Cd,Co)S,\cite{Gennser:1995} a value
close to those of Mn-based DMS.\cite{Dietl:1994} Similar
magnitudes of $\alpha N_0$ in (II,Co)VI and (II,Mn)VI compounds
is consistent with the virtually identical values of the
the s-d exchange energy for the free Co$^{+1}$ and Mn$^{+1}$ ions,
$J_{\mathrm{s-d}} = 0.40$~eV.\cite{Landolt:1950}
The temperature and field dependent mean spin value along the
field direction $\langle S_z\rangle$ is numerically calculated
from the spin hamiltonian of Co$^{2+}$ ions in ZnO.\cite{Sati:2006}
 We assume here that a half of Co ions is
randomly distributed over the paramagnetic host, which leads to
$x_{\mathrm{eff}} = 0.018$ for $x = 0.025$.\cite{Dietl:1994}

Figures~\ref{fig:MRZnCoO}(c,d) show the computed values of MR. In view
that no adjustable parameters are involved, the agreement between
experimental [Figs.~\ref{fig:MRZnCoO}(a,b)] and theoretical
[Figs.~\ref{fig:MRZnCoO}(c,d)] findings is to be regarded as quite good.
In particular, the theory reproduces MR shape and temperature
dependence. Furthermore, clearly visible MR differences for the
two configurations, $H\parallel c$ and $H\perp c$, are well
reproduced by the theory.

\section{Conclusions}

The findings presented in this paper together with the results
accumulated over recent years for (Zn,Co)O can consistently be interpreted in terms of spinodal decomposition into Co-rich antiferromagnetic nanocrystals
embedded in the Co-poor (Zn,Co)O host. Within this scenario, presumably relevant to a number of other DMS and DMO, ferromagnetic signatures in the absence
of ferromagnetic precipitates such as free standing Co clusters,
result from uncompensated spins at the surface of the antiferromagnetic nanocrystals.
The nanocrystal aggregation can be steered by growth conditions and co-doping,\cite{Dietl:2006_a,Kuroda:2007} as the position of the Fermi level affects the charge state and/or and the diffusion coefficient of transition metals in semiconductors.  This explains the sensitivity of the ferromagnetic response to co-doping and deviations from stoichiometry found in (Zn,Co)O.\cite{Rode:2003,Venkatesan:2004,Chambers:2006} Importantly, this means that self-organized nano-assembling of magnetic nanocrystals in a semiconductor matrix
can be controlled during the growth process.

\begin{acknowledgements}
The work in Warsaw was supported in part by ERATO Semiconductor Spintronics Project of Japan Science
and Technology Agency and by SPINTRA Project of European Science Foundation.  The work at
the National University of Singapore was supported by the A*-STAR under Grant No.~R-398-000-020-305.
The authors are grateful to Zaibing Guo for the SQUID measurement.
\end{acknowledgements}


\end{document}